# STUDY ON HIGH ORDER MODES IN A BETA=0.09 SRF HWR CAVITY FOR 100 mA PROTON ACCELERATION[*]

H.T.X. ZHONG, F. Zhu[#], S.W. Quan, F. Wang, K.X. Liu

Institute of Heavy Ion Physics & State Key Laboratory of Nuclear Physics and Technology, Peking University, Beijing, 100871, China

**Abstract**: There's presently a growing demand for cw high current proton and deuteron linear accelerators based on superconducting technology to better support various fields of science. Up to now, high order modes (HOMs) studies induced by ion beams with current higher than 10 mA and even 100 mA accelerated by low β non-elliptical Superconducting rf (SRF) cavities are very few. Peking University has recently designed and fabricated a β=0.09 162.5 MHz HWR cavity to study the key physics problems in accelerating beam with current of about 100mA. This paper focuses on the study of the HOM-induced power in this cavity. The incoherent beam energy loss induced by 100 mA beam through the HWR SRF cavity were obtained from both time domain solver and frequency domain eigenmodes spectrum method. We also analysed the possibility of coherent excitation of this cavity considering of manufacture errors.

## INTRODUCTION

Compared to normal conducting accelerator, rf Superconducting accelerator has more advantages and the potential to accelerate super high current (for example 100 mA) cw ion beam. The beam pipes can be larger and the operation cost could be much less. Such high current SRF cavities have been adopted by some future facilities. For example, IFMIF has two 125 mA deuteron accelerators [1] and BISOL [2] proposed a 50 mA deuteron accelerator as a driver. Supported by National Basic Research Project, Peking University (PKU) is developing a β =0.09 HWR SRF prototype cavity for 100mA proton beam acceleration.

Compared to elliptical SRF cavities, quarter wave resonators (QWRs) or half wave resonators (HWRs) have much sparse high order modes. The modes are a little far from the accelerating mode and not easily activated. At low current case, the effect of the HOMs of QWRs or HWRs can be negligible and the studies of HOMs of HWRs are very few. But for 100 mA beam, whether the beam energy loss induced by the HOMs can be negligible or not still needs study.

HOMs contribute to the additional cryogenic load. Cavity loss factor calculation is very important for the total cryo-losses estimation for the SRF cavities. In this paper, we will describe our efforts to characterize the beam-induced HOM power in Peking University's β=0.09 HWR cavity for 100 mA beam acceleration, and present the results of incoherent losses calculations (in time domain and frequency domain) and the possibility of coherent excitation.

## PROPERTIES OF THE HWR CAVITY

The β=0.09 162.5 MHz HWR cavity fabricated by Peking University is supposed to accelerate 100 mA proton beam or 50 mA deuteron beam after RFQ. It has taper conductors to achieve low peak magnetic field and good mechanical stability. Figure 1 shows the structure of the cavity.

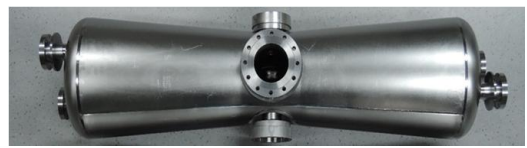

Figure 1: The 162.5 MHz, high current β=0.09 HWR cavity.

Table 1 gives the main geometry and RF parameters of the HWR cavity. Since the low design beta and high beam intensity, the cavity has a slim body and comparative big aperture. The short plates are specially designed with combined plain part and round part to suppress multipacting. Mechanical simulation results gives that the cavity has very low df/dP and Lorentz froce coefficient[3]. This taper type HWR cavity has better mechanical propertiesthan cylindrical HWR cavity[4].

[*] Work supported by National Basic Research Project (No.2014CB845504)
[#] zhufeng7726@pku.edu.cn

Table 1: RF and geometry parameters of the HWR cavity

| Parameter | Value |
|---|---|
| Frequency/MHz | 162.5 |
| Optimal β | 0.09 |
| Cavity diameter /mm | 260 |
| Beam aperture /mm | 40 |
| Cavity height /mm | 990 |
| $L_{cav}=\beta\lambda$ /mm | 166 |
| R/Q /Ω | 255 |
| Geometry factor /Ω | 39 |
| Bpk/Eacc /(mT/(MV/m)) | 6.4 |
| Epk/Eacc | 5.3 |

## ROOM TEMPERATURE MEASUREMENTS

Two prototype HWR cavities have already been fabricated. We have done some room temperature measurements on one of the HWRs to get experimental data of the HOMs. Figure 2 shows the layout of the experiment system of the HWR cavity.

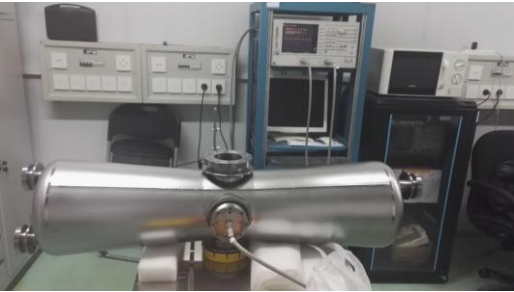

Figure 2: HWR009 cavity RF measurements at room temperature.

First, we used a vector network analyzer to measure the frequency and $Q_L$ of the cavity's monopole eigenmodes. Figure 3 shows the measured HOM spectrum. Table 2 gives the measured $Q_0$ and frequencies of HOMs (both experiment and simulation results). Figure 4 shows some calculated monopole modes with relatively high R/Q of the HWR cavity. The upper frequency of the HOMs we studied in this paper is 1.3 GHz. We think the limited upper frequency is not a problem since modes with high frequency we calculated all have a very small R/Q (lower than mΩ).

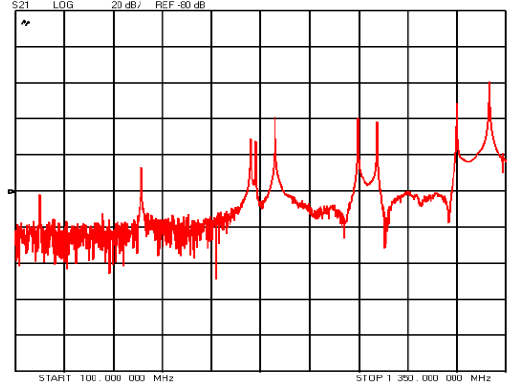

Figure 3: HOM spectrum of the HWR cavity

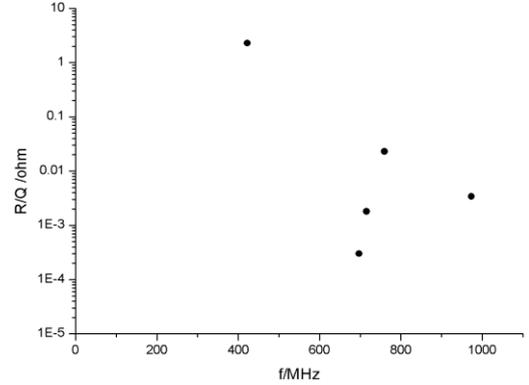

Figure 4: Some calculated monopole modes with relatively high R/Q of the HWR cavity.

Table 2: RF and geometry parameters of the HWR cavity

| Mode | Freq(simulation)/MHz | Freq(experiment)/MHz | $Q_0$ |
|---|---|---|---|
| 1 | 162.23 | 162.22 | 3900 |
| 2 | 422.35 | 420.78 | 5573 |
| 3 | 697.35 | 699.06 | 1338 |
| 4 | 715.51 | 713.22 | 6934 |
| 5 | 759.86 | 761.87 | 6070 |
| 6 | 973.25 | 972.97 | 1006 |
| 7 | 1004.43 | 1002.09 | - |
| 8 | 1160.36- | 1157.68 | 1109 |
| 9 | 1226.76- | 1225.68 | 1498 |
| 10 | 1232.05- | 1234.05 | 1004 |
| 11 | 1243.51 | 1240.25 | 9892 |

We also used a perturbation method to measure the axial electric field in this cavity with a small aluminium ball. The experimental R/Q is about 278Ω, which is similar as the simulated one of 255Ω. We can say that the experimental results and simulative results are in good agreement.

## TIME DOMAIN ANALYSIS

The power deposited by the beam consisting of bunches passing through the cavity with the bunch repetition rate $f_{rep}$ is

$$P = k_{//}Iq = k_{//}I^2/f_{rep} \qquad (1)$$

Where $I=qf_{rep}$ is the average beam current, q is the bunch charge, and $k_{//}$ is the beam energy loss factor.

The time domain calculation of beam energy loss factors is very common and well developed for elliptical cavities and for relativistic beams. Code ABCI can calculate the loss factor of symmetric structure and for relativistic beams [5]. CST can calculate the loss factor of 3D structure, but normally also for relativistic beams. When simulating non-relativistic beams passing through a cavity, one needs to take into account the static Coulomb forces. CST Studio direct wake field solver was used to calculate wake potentials. The total wake potential includes both static (Coulomb forces) and dynamic (beam-cavity interaction) parts. Because of numerical noise present in the direct solution the static component is not perfectly symmetrical to the bunch center, and thus the convolution of the bunch profile with the wake potential gives the wrong result for the loss factor. The remedy is to run two consecutive simulations with slightly different pipe lengths, and then the static components of the wake potential will change proportionally to the length while the dynamic part remains the same [6]. Thus, from these two solutions it is possible to subtract the static part and find the wake potential caused by beam-cavity interactions only. We used this method to calculate the wake potentials and further the loss factor for the high current taper type β=0.09 HWR cavity.

CST calculation requires a meshing of the full structure volume, and the total number of mesh elements exceeds ten million for short bunches. The shorter the bunch length is, the more calculation time consumes. The beam bunch which we were able to simulate within a reasonable time was 3 mm rms length.

We got the dynamic wake potentials from cavities with different beam pipe lengths. The loss factor was then obtained through the convolution of the bunch profile with the dynamic wake potential. Figure 5 shows the loss factor changes as the beam velocity. On one hand, when β is smaller than 0.09 the loss factor is relative low. On the other hand, there is no much difference of the loss factors when the ion beam velocity β is in the range of 0.09~0.14.

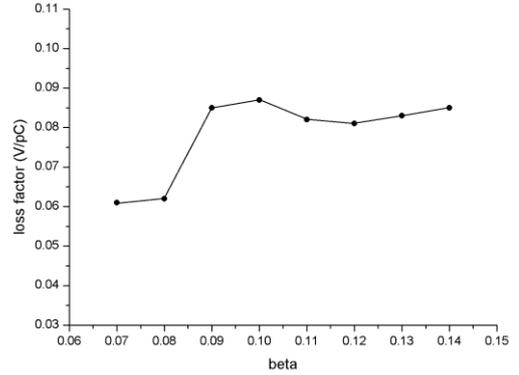

Figure 5: loss factor versus velocity β for the HWR cavity.

If we choose $k_{//}$ =0.085 V/pC for the β =0.09 HWR cavity, the average beam current is 100 mA, and the bunch repetition rate is 162.5 MHz, then the incoherent bunch energy loss is about 5.2 W from equation (1).

## FREQUENCY DOMAIN ANALYSIS

The total loss factor k can be represented as an infinite series of all cavity modes' inputs. The loss factor for an individual mode for a Gaussian bunch with rms length σ can be written in the form [7]:

$$k(\beta,\sigma) = \exp\left[-\left(\frac{\omega\sigma}{\beta c}\right)^2\right]\frac{\omega r(\beta)}{4Q} \qquad (2)$$

where ω is the circular eigenfrequency, r(β) is the shunt impedance, Q is the eigenmode quality factor and βc is the bunch velocity.

In superconducting cavities, the contribution of the lowest monopole modes is a major concern. Meanwhile, modes' R/Q depend on bunch's velocity and length. Figure 6 shows the 3 mm Gaussian bunch's incoherent HOM losses at different beam velocities. Those losses do not include modes with frequencies higher than 1.3 GHz.

From Figure 6, we can conclude that the total loss factor of the HWR cavity for β=0.09 particles from the frequency domain is about 0.067 V/pC. The majority is from the fundamental mode and the contribution of all the other modes gives 0.0015 V/pC. The highest loss factor of all the HOMs of this cavity is 0.0126 V/pC when the velocity is β =0.14. We find that the loss factor of HOMs is small and most of the loss comes from the mode with frequency of 422.28 MHz . If the average beam current is 100 mA, and the bunch repetition rate is 162.5 MHz, then the incoherent bunch energy loss from frequency domain analysis is about 4.1 W. The energy loss caused by HOMs is 0.77 W for β =0.14 particles and is smaller for lower velocity beam.

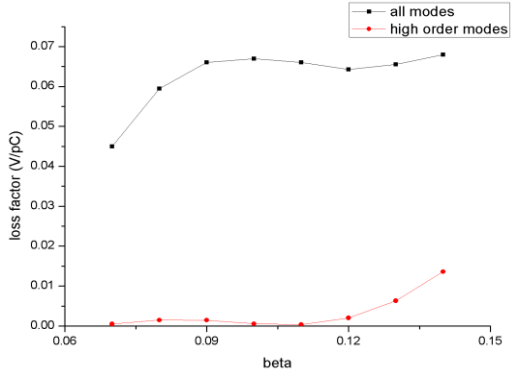

Figure 6: Loss factor versus velocity β for the HWR cavity. The black line is the loss factor of all modes. The red line is the loss factor of high order modes (except the accelerating mode)

By comparing the frequency domain analysis and the time domain analysis, we find that the loss factor results have little difference, but the relationships between loss factor and beam velocity have similar trend. We think the frequency domain results are more accurate, because the static wake potentials are much higher than the dynamic wake potentials in such a low beta HWR cavity given by the CST code and thus bigger errors exist in time domain calculations.

## RESONANCE HOM EXCITATION

Resonance HOM excitation can happen when a bunched continuous beam passes through a superconducting cavity whose repetition frequency's integral multiple is very close to one of the HOM's frequencies. In this situation, the HOM losses can grow up to hundreds of times.

$$U_{HOM} = \frac{\tilde{I}(R/Q)}{4\delta f/f} \quad (3)$$

$$P_{loss} = \frac{U_{HOM}^2}{(\frac{R}{Q})Q_0} \quad (4)$$

We can use equation (3),(4) as a simplified way to calculate a mode's losses of possible coherent excitation[8]. $\delta f$ is the difference between the mode's frequency and the integral multiple of bunch's repetition frequency.

In the β=0.09 HWR cavity, most of modes' $\delta f$ are larger than 10 MHz when operating at 162.5 MHz CW mode. Only the 15th mode (frequency 973.15 MHz) has the smallest $\delta f$=1.85 MHz.

In order to accurately estimate the probability of resonance HOM excitation, we need to figure out the fluctuation of HOM frequency as well. Since it was not available for us to do this work with experiment and statistic analysis, we did HOM simulations taking into account manufacturing mechanical tolerances instead. In every individual simulation, we randomly varied the cavity's structure within ±0.2 mm tolerances around standard values. Through these simulations we found the limit of the frequency shift was about 1.4 MHz, a small value compared to elliptical cavities. After that, we calculate the HOM power in the worst situation where the mode's frequency is 974.55 MHz and current is 100 mA. In this situation, the mode's loss is about $164/Q_0$ W($Q_0$ is cavity's quality factor), a very small value for SRF cavities. So it seems that the resonance excitation is not a problem for this HWR cavity when it operates at 162.5 MHz CW mode.

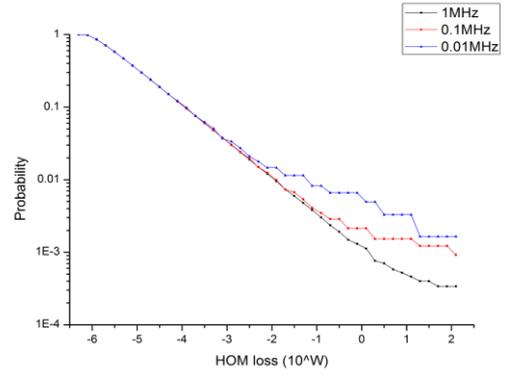

Figure 7: Probability distribution of HOM losses. The point presents the probability of cavity having a HOM losses over a specific power.

We also studied the HOM resonance probabilities at lower operating repetition rates which might be adopted in some user applications. In this situation, the beam current is not high as 100 mA, but the resonance condition is easier to be met. Assuming high order modes' frequency conforms to normal distribution with a variance of 0.5 MHz and all modes have same $Q_l=Q_0=2\times 10^9$. Then we calculated the HOM power ten millions time with randomly HOM frequency and bunch charge of 1nC to get the probability distribution of HOM power losses at different bunch repetition rates. Figure 7 shows the result, the rise at the end of curves is a result of intense coherent excitation. The probability of a large HOM power over 10 W is below $2\times 10^{-4}$ when the repetition rate is higher than 1 MHz and could be $2\times 10^{-3}$ when the repetition rate is 10 kHz. So the cavity is not likely to have a harmful coherent excitation. Even if happened, we can shift the HOM frequency by detuning cavity's accelerating mode by few tens of kHz and then tuning operating mode back to resonance.

## CONCLUSION

The estimation of the total beam energy loss of 100 mA beam passing through a β=0.09 HWR cavity was made. We use two independent methods (in time and frequency domains) for incoherent loss factor calculation. The final amount of incoherent RF losses calculated for a single cavity is about 4.1 W while the energy loss induced by the HOMs is only a small part of it. Meanwhile we have considered the possible coherent excitation of the HOM spectrum. caused by manufacturing errors, or in several situations. The resonance excitation can barely happen when cavity operates at 162.5 MHz CW mode and the probabilities of a high HOM power when cavity is operating at other repetition rates are not high.